\documentclass[12pt]{article}

\usepackage{amsmath,amssymb}
\usepackage{epic,eepic}

\DeclareMathOperator{\tr}{tr}
\DeclareMathOperator{\Tr}{Tr}
\DeclareMathOperator{\dd}{d}

\newcommand{\MM}{M}

\newcommand{\GG}{G}

\newcommand{\EE}{E}

\newcommand{\CCC}{\mathbb{C}}

\newcommand{\pr} {\partial}

\newcommand{\AAA}{\mathcal{S}}

\newcommand{\ff}{\mathcal{A}}

\newcommand{\JJ} {\mathcal{G}}

\newcommand{\HH} {\mathcal{H}}

\newcommand{\Length}{\text{L}}
\newcommand{\Mass}{\text{M}}
\newcommand{\Time}{\text{T}}

\begin{document}

\begin{center}
{
\bf \large Mass in Quantum Yang-Mills Theory  \\
}
(Comment on a Clay Millenium Problem)\\

\vskip 0.3cm

{\bf L.~D.~Faddeev } \\
St.~Petersburg Department of Steklov \\
Mathematical Institute.
\end{center}

	Among seven problems, proposed for XXI century by Clay 
	Mathematical Institute
\cite{clay},
	there are two stemming from physics. One of them is 
	called 
	"Yang-Mills Existence and Mass Gap".
	The detailed statement of the problem, written by
	A.~Jaffe and E.~Witten
\cite{JW},
	gives both motivation and exposition of related mathematical
	results, known until now.
	Having some experience in the matter, I decided to 
	completement their text	by my own personal 
comments\footnote{The first variant was published in 
\cite{comment}.
In this new version more details are given in the description
of renormalization.} 
	aimed mostly to mathematical audience.

\section{What is Yang-Mills field}
	Yang-Mills field bears the name of the authors of the
	famous paper
\cite{YM},
	in which it was introduced into physics. From mathematical
	point of view it is a connection in a fiber bundle with
	compact group
    $ \GG $
	as a structure group. We shall treat the case when the 
	corresponding principal bundle
    $ \EE $
	is trivial
\begin{equation*}
	\EE = \MM_{4} \times \GG
\end{equation*}
	and the base
    $ \MM_{4} $
	is a four dimensional Minkowski space.

	In our setting it is convenient to describe the Yang-Mills 
	field as one-form
    $ A $
	on
    $ \MM_{4} $
	with the values in the Lie algebra
    $ \JJ $
	of
    $ \GG $:
\begin{equation*}
        A(x) = A^{a}_{\mu} (x) t^{a} dx^{\mu}.
\end{equation*}
	Here
    $ x^{\mu} $, 
    $ \mu = 0,1,2,3 $ 
	are coordinates on
    $ \MM_{4} $;
    $ t^{a} $,
    $ a = 1, \ldots , \dim G $ ---
	basis of generators of
    $ \JJ $
	and we use the traditional convention of taking sum over
	indices entering twice.

	Local rotation of the frame
\begin{equation*}
        t^{a} \to h(x) t^{a} h^{-1}(x) ,
\end{equation*}
	where
    $ h(x) $
	is a function on
    $ \MM_{4} $
	with the values in
    $ \GG $
	induces the transformation of the
    $ A $
	(gauge transformation)
\begin{equation*}
        A(x) \to h^{-1}(x) A(x) h(x) + h^{-1} dh(x) = A^{h}(x) .
\end{equation*}
	Important equivalence principle states, that a physical
	configuration is not a given field
    $ A $,
	but rather a class of gauge equivalent fields. This principle
	essentially uniquely defines the dynamics of the
	Yang-Mills field.

	Indeed, the action functional, leading to the equation of motion via
	variational principle, must be gauge invariant. Only one local
	functional of second order in derivatives of
    $ A $
	can be constructed.

	For that we introduce the curvature --- two form with values in
    $ \JJ $
\begin{equation*}
	F = d A + A^{2} ,
\end{equation*}
	where the second term in RHS is exterior product of one-form
	and commutator in
    $ \JJ $.
	In more detail
\begin{equation*}
        F = F^{a}_{\mu\nu} t^{a} dx^{\mu} \wedge dx^{\nu} ,
\end{equation*}
	where
\begin{equation*}
        F^{a}_{\mu\nu} = \pr_{\mu} A_{\nu} - \pr_{\nu} A_{\mu} +
		f^{abc} A^{b}_{\mu} A^{c}_{\nu}
\end{equation*}
	and
    $ f^{abc} $
	are structure constants of
    $ \JJ $
	entering the basic commutation relation
\begin{equation*}
        [ t^{a} , t^{b} ] = f^{abc} t^{c} .
\end{equation*}
	The gauge transformation of
    $ F $
	is homogenous
\begin{equation*}
        F \to h^{-1} F h ,
\end{equation*}
	so that the 4-form
\begin{equation*}
        \ff = \tr F \wedge F^{*} = F^{a}_{\mu\nu} F^{a}_{\mu\nu} d^{4}x
\end{equation*}
	is gauge invariant. Here
    $ F^{*} $
	is a Hodge dual to
    $ F $
	with respect to Minkowskian metric, and
    $ d^{4}x $
	is corresponding volume element. It is clear, that
    $ S $
	contains the derivatives of
    $ A $
	at most in second order. The integral
\begin{equation}
\label{AC}
        \AAA = \frac{1}{4 g^{2}} \int_{\MM_{4}} \ff
\end{equation}
	can be taken as an action functional. The positive constant
    $ g^{2} $
	in front of the integral is a dimensionless parameter which
	is called a coupling constant. Let us stress, that it is
	dimensionless only in the case of four dimensional space-time.

	Remind that in general the dimension of physical quantity
	is a product of powers of 3 fundamental dimensions
    $ [ \Length ] $ ---
	length,
    $ [ \Time ] $ ---
	time and
    $ [ \Mass ] $ ---
	mass with usual units of cm, sec and gr. However in
	relativistic quantum physics we have two fundamental constants ---
	velocity of light
    $ c $
	and Planck constant
    $ \hbar $
	and use the convention, that
    $ c = 1 $ and
    $ \hbar = 1 $,
	reducing the possible dimensions to the powers of lenght
    $ [ \Length ] $.
	The Yang-Mills field has dimension
    $ [ A ] = [ \Length ]^{-1} $,
	the curvature
    $ [ F ] = [ \Length ]^{-2} $,
	the volume element
    $ [ d^{4}x ] = [ \Length ]^{4} $,
	so that integral in
    $ \AAA $
	is dimensionless.
	Now full
    $ \AAA $
	should be dimensionless, as it has the same dimension as
    $ \hbar $,
	thus 
    $ g^{2} $
	has dimension zero.

	We see, that
    $ \AAA $
	contains terms in powers of
    $ A $
	of degrees 2, 3, 4
\begin{equation*}
        \AAA = \AAA_{2} + \AAA_{3} + \AAA_{4} ,
\end{equation*}
	which means that Yang-Mills field is selfinteracting.

	Among many approaches to quantizing the Yang-Mills theory
	the most natural is that of the functional integral. 
	Indeed, the equivalence principle is taken into account in
	this approach by integrating over classes of equivalent
	fields. 
	So we shall use this approach in what follows.
	There is no place here to describe in detail this purely
	heuristic method
	of quantization, moreover it hardly will lead to a solution
	of Clay Problem. 
	However it will be very usefull for an intuitive explanation
	of this problem, which we shall do here.

\section{What is mass}
	It was the advent of the special relativity which has given a
	natural definition of mass. A free massive particle has 
	the following expression of the energy
    $ \omega $
	in terms of its momentum
    $ p $
\begin{equation*}
	\omega(p) = \sqrt{p^{2} + m^{2}} ,
\end{equation*}
	where
    $ m $
	is called mass.
	In quantum version mass appears as a parameter (one out of two)
	of the irreducible representation of the Poincare group
	(group of motion of the Minkowski space).

	In quantum field theory this representation 
	(insofar as
    $ m $) 
	defines a one-particle space of states
    $ \HH_{m} $
	for a particular particle entering the full spectrum of particles.
	The state vectors in such a space can be described as
	functions
    $ \psi(p) $
	of momentum
    $ p $
	and
    $ \omega(p) $
	defines the energy operator.

	The full space of states has the structure
\begin{equation*}
        \HH = \CCC \oplus \left( \sum_{i} \oplus \HH_{m_{i}} \right)
		\oplus \cdots ,
\end{equation*}
	where one dimensional space 
    $ \CCC $
	corresponds to the vacuum state
	and 
    $ \cdots $ 
	mean spaces of many-particles states, being tensor products 
	of one-partical	spaces. 
	In particular if all particles in the system are massive
	the energy has zero eigenvalue corresponding to vacuum and then
	positive continuous spectrum from
    $ \min m_{k} $
	till infinity. In other words the least mass defines the gap in
	the spectrum. The Clay problem requires the proof of such a gap
	for the Yang-Mills theory.

	We see an immediate difficulty. In our units
    $ m $
	has dimension
    $ [m] = [ \Length ]^{-1} $.
	But in the formulation of the 
	classical Yang-Mills theory no dimesional parameter entered.
	On the other hand, the Clay Problem requires, that in quantum
	version such parameter must appear. How come?

	I decided to write these comments exactly for the explanation 
	how quantization can lead to appearence of dimensional parameter
	when classical theory does not have it. This possibility is
	connected with the fact, that quantization of the interacting
	relativistic field theories leads to infinities --- appearence
	of the divergent integrals which are dealt with by the proccess
	of renormalization. Traditionally these infinities were 
	considered as a plague of the Quantum Field Theory.
	One can find very strong words denouncing them, belonging to the
	great figures of several generations, such like Dirac, Feynmann
	and others. However I shall try to show, that the infinities in
	the Yang-Mills theory are beneficial --- they lead to
	appearence of the dimensional parameter after the quantization
	of this theory.

	This point of view was already emphasized by R.~Jackiw
\cite{RJ}
	but to my knowledge it is not shared yet by other specialists.

	Sidney Coleman
\cite{SC}
	coined a nice name "dimensional transmutation" for the phenomenon, 
	which I am going to describe. Let us see what all this
	means.

\section{Dimensional transmutation}
	The most direct way to introduce the functional integral
	is to consider the generating functional for the scattering
	operator.
	This functional depends on the initial and final configuration
	of fields, defined by the appropriate asymptotic condition.
	In naive formulation these asymptotic configurations are
	given as solutions
    $ A_{\text{in}} $ and
    $ A_{\text{out}} $
	of the linearized classical equations of motion.
	Through these solutions the particle interpretation is introduced
	via well defined quantization of the free fields.
	However the more thorough approach leads to the corrections,
	which take into account the selfinteraction of particles.
	We shall see below, how it is realized in some consistent
	way.

	Very formally the generating functional
    $ W(A_{\text{in}},A_{\text{out}}) $
	is introduced as follows
\begin{equation}
\label{genfun}
    e^{i W(A_{\text{in}}, A_{\text{out}})} = \int_{A \to 
	    \begin{subarray}{l} 
		A_{\text{in}}, t \to - \infty \\
		A_{\text{out}}, t \to + \infty
	    \end{subarray}}
    e^{i S(A)} d A ,
\end{equation}
	where
    $ S(A) $
	is the classical action
(\ref{AC}).
	Symbol
    $ d A $
	denotes the integration measure and we shall make it more 
	explicit momentarily.

	The only functional integral one can deal with is a gaussian one.
	To reduce
(\ref{genfun})
	to this form and, in particular, to identify corresponding quadratic
	form we make shift of integration variable
\begin{equation*}
    A = B + g a ,
\end{equation*}
	where the external variable
    $ B $ 
	should take into account the asymptotic boundary conditions 
	and new integration variable 
    $ a $
	has zero incoming and outgoing components.

	We can consider both
    $ A $
	and
    $ B $
	as connections, then 
    $ a $
	will have only homogeneous gauge transformation
\begin{equation*}
    a(x) \to h^{-1}(x) a h(x) .
\end{equation*}
	However, for fixed 
    $ B $
	the transformation law for
    $ a $
	is nonhomogeneous
\begin{equation}
\label{go}
    a \to a^{h} = \frac{1}{g} (A^{h} - B) .
\end{equation}
	Thus the functional
    $ S(B+a) - S(B)$
	is constant along such ``gauge orbits''.
	Integration over
    $ a $
	is to take this into account.
	We shall denote
    $ W(A_{\text{in}}, A_{\text{out}}) $
	as
    $ W(B) $,
	having in mind that 
    $ B $
	is defined by
    $ A_{\text{in}} $,
    $ A_{\text{out}} $
	via some differential equation.
	Here is the answer detailing the formula
(\ref{genfun})
\begin{equation}
\label{idet}
\begin{split}
    e^{i W(B)} = e^{i S(B)} \int \exp i	\bigl\{ S(B&+a)-S(B) 
	    + \int \frac{1}{2} \tr (\nabla_{\mu} a_{\mu})^{2}dx\bigr\} \\
	& \times \det \bigl((\nabla_{\mu}+ g a_{\mu}) 
	    \nabla_{\mu}\bigr) \prod_{x} d a(x) .
\end{split}
\end{equation}
	Here we integrate over all variables
    $ a(x) $,
	considered as independent coordinates.
	Furthermore,
    $ \nabla_{\mu} $
	is a covariant derivative with respect to connection
    $ B $
\begin{equation*}
    \nabla_{\mu} = \partial_{\mu} + B_{\mu} .
\end{equation*}
	The quadratic form
    $ \frac{1}{2} \int (\nabla_{\mu} a_{\mu})^{2} dx $
	regularizes the integration along the gauge orbits
(\ref{go})
	and the determinant provides the appropriate normalization.
	This normalization was first realized by V.~Popov and me
\cite{FP}
	with additional clarification by 't~Hooft
\cite{Hooft}.
	I refer to physical literature
\cite{FS}, \cite{Peskin}
	for all explanations.
	One more trick consists in writing the determinant in terms 
	of the functional integral
\begin{equation*}
    \det (\nabla_{\mu} + g a_{\mu}) \nabla_{\mu} = \int
	\exp i \bigl\{ \int \tr \bigl( (\nabla_{\mu} + g a_{\mu})
	    \bar{c} \nabla_{\mu} c \bigr) dx \bigr\}
	\prod_{x} d \bar{c}(x) d c(x)
\end{equation*}
	over grassman algebra with generators
    $ \bar{c}(x) $, $ c(x) $
	in the sense of Berezin
\cite{Berezin}.
	These anticommuting field variables play only accessory role,
	there are no physical degrees of freedom, corresponding
	to them.

	The resulting functional which we should integrate over
    $ a(x) $, $ \bar{c}(x) $, $ c(x) $
	assumes the form
\begin{equation}
\begin{split}
    \exp i \Bigl\{ \frac{1}{2} (M_{1}a, a) & +(M_{0} \bar{c}, c)
	+ \frac{1}{g} \Gamma_{1}(a)  \\
	  &+ g \Gamma_{3}(a,a,a) + g^{2} \Gamma_{4}(a,a,a,a)
	    + g \Omega_{3} (\bar{c}, c, a) \Bigr\} ,
\end{split}
\end{equation}
	where we use short notations for the corresponding linear,
	quadratic, cubic and quartic forms in variables
    $ a $ and
    $ \bar{c} $, $ c $.
	The linear form
    $ \Gamma_{1}(a) $
	is defined via the classical equation of motion for
	the field
    $ B_{\mu}(x) $
\begin{equation}
\label{G1}
    \Gamma_{1}(a) = \int \tr (\nabla_{\mu} F_{\mu\nu}(x) a_{\nu}(x)) dx ,
\end{equation}
	forms 
    $ \Gamma_{3} $,
    $ \Gamma_{4} $
	and
    $ \Omega_{3} $
	are given by
\begin{align}
    \Gamma_{3} & = \int \tr \nabla_{\mu} a_{\nu} [a_{\mu}, a_{\nu}]
	dx ,\\
    \Gamma_{4} & = \frac{1}{4} \int \tr \bigl([a_{\mu}, a_{\nu}]
	\bigr)^{2} dx ,\\
\label{O3}
    \Omega_{3} & = \int \tr \nabla_{\mu} \bar{c} [a_{\mu}, c
] dx
\end{align}
	and operators 
    $ M_{1} $ and
    $ M_{0} $
	of the quadratic forms look like
\begin{align}
    M_{1} & = - \nabla_{\rho}^{2} \delta_{\mu\nu} - 2 [F_{\mu\nu}, \cdot] ,\\
    M_{0} & = - \nabla_{\rho}^{2} .
\end{align}
	The equation on the external field
    $ B $
	in the naive approach would be classical equation of motion,
	assuring vanishing of 
    $ \Gamma_{1}(a) $.
	This would correspond to the stationary phase method.
	However we shall make a different choice taking into account
	the appropriate quantum corrections.

	It is instructive to use the simple pictures (Feynman diagrams)
	to visualize the objects
(\ref{G1})--(\ref{O3}).
	For the forms
    $ \Gamma_{1} $,
    $ \Gamma_{3} $,
    $ \Gamma_{4} $
	and
    $ \Omega_{3} $
	they look as vertices with external lines, number of which
	equals number of fields
    $ a(x) $, $ \bar{c}(x) $, $ c(x) $
\begin{align}
\label{diag1}
\begin{picture}(40,30)
    \drawline(5,15)(35,15)
    \drawline(1,17)(5,13)
    \drawline(1,13)(5,17)
\end{picture} & &
\begin{picture}(40,30)
    \drawline(20,14)(20,29)
    \drawline(20,14)(9,1)
    \drawline(20,14)(31,1)
\end{picture} & &
\begin{picture}(40,30)
    \drawline(6,1)(34,29)
    \drawline(6,29)(34,1)
\end{picture} & &
\begin{picture}(40,30)
    \dashline{3}(4,1)(18,15)(4,29)
    \drawline(19,15)(33,15)
    \drawline(9,8)(12,9)(11,6)
    \drawline(13,22)(10,23)(11,20)
\end{picture}
\\
\nonumber
    \Gamma_{1} & & \Gamma_{3} & & \Gamma_{4} & &\Omega_{3}
\end{align}
	The Green functions
    $ G_{1} $ and
    $ G_{0} $
	for operators
    $ M_{1} $ and 
    $ M_{0} $
	are depicted as simple lines
\begin{align}
\label{diag2}
\begin{picture}(40,20)
    \drawline(5,10)(35,10)
\end{picture} & &
\begin{picture}(40,20)
    \dashline{4}(6,10)(34,10)
    \drawline(17,12)(21,10)(17,8)
\end{picture} & &
\\
\nonumber
    G_{1} & & G_{0} & &
\end{align}
	Each end of lines in
(\ref{diag1}) and
(\ref{diag2})
	bears indices
    $ x, \mu, a $ or
    $ x, a $
	characterizing fields
    $ a_{\mu}^{a}(x) $ and
    $ \bar{c}^{a}(x) $, 
    $ c^{b}(x) $.
	The arrow on line
\begin{picture}(30,8)
    \dashline{4}(1,4)(29,4)
    \drawline(12,6)(16,4)(12,2)
\end{picture}
	distinguishes fields 
    $ \bar{c} $ and
    $ c $.
	Note, that Green functions are well defined due to homogeneous
	boundary conditions for
    $ a(x) $, $ \bar{c}(x) $, $ c(x) $.

	Now simple combinatorics for the gaussian integral which we get from
(\ref{idet})
	expanding the exponent, containing vertices, in a formal series,
	gives the following answer
\begin{equation}
\begin{split}
    \exp i W(B) = &\exp i S(B) (\det M_{1})^{-1/2}
	\det M_{0} \\
    & \times \exp \{ \sum \text{connected closed graphs}\} ,
\end{split}
\end{equation}
	where we get graph by saturating the ends of vertices
	by lines, corresponding to the Green functions.
	The term ``closed'' means, that graph have no external lines.

	We shall distinguish weakly and strongly connected graphs.
	The weakly connected graph can be made disconnected by 
	crossing one line. 
	(In physical literature the term ``one	particle reducible'' 
	is used for such graph.)

	The quantum equation of motion, which we impose on
    $ B $,
	can be depicted as
\begin{equation}
\label{em}
\begin{picture}(50,20)(0,10)
    \drawline(5,15)(35,15)
    \drawline(1,17)(5,13)
    \drawline(1,13)(5,17)
\end{picture} +
\begin{picture}(60,20)(-10,10)
    \drawline(21,15)(39,15)
\texture{55888888 88555555 5522a222 a2555555 55888888 88555555 552a2a2a 2a555555
    55888888 88555555 55a222a2 22555555 55888888 88555555 552a2a2a 2a555555
    55888888 88555555 5522a222 a2555555 55888888 88555555 552a2a2a 2a555555
    55888888 88555555 55a222a2 22555555 55888888 88555555 552a2a2a 2a555555
}
    \put(10,15){\shade\ellipse{20}{20}}
\end{picture} = 0 ,
\end{equation}
	where the second term in the LHS is a sum of strongly
	connected graphs with one external line.
	In the lowest approximation it looks as follows
\begin{equation*}
\begin{picture}(45,20)(0,10)
    \drawline(5,15)(35,15)
    \drawline(1,17)(5,13)
    \drawline(1,13)(5,17)
\end{picture} + g
\begin{picture}(55,20)(-5,10)
    \drawline(21,15)(39,15)
    \put(10,15){\ellipse{20}{20}}
\end{picture} + g
\begin{picture}(55,20)(-5,10)
    \drawline(21,15)(39,15)
    \put(10,15){\arc{20}{0}{0.3}}
    \put(10,15){\arc{20}{0.6}{0.9}}
    \put(10,15){\arc{20}{1.2}{1.5}}
    \put(10,15){\arc{20}{1.8}{2.1}}
    \put(10,15){\arc{20}{2.4}{2.7}}
    \put(10,15){\arc{20}{3.0}{3.3}}
    \put(10,15){\arc{20}{3.6}{3.9}}
    \put(10,15){\arc{20}{4.2}{4.5}}
    \put(10,15){\arc{20}{4.8}{5.1}}
    \put(10,15){\arc{20}{5.4}{5.7}}
    \put(10,15){\arc{20}{6.0}{6.3}}
    \drawline(12,27)(9,25)(12,23)
\end{picture} 
    = 0 .
\end{equation*}
	With this understanding the expression for
    $ W(B) $
	is given by the series in the powers of the coupling constant
    $ g^{2} $
\begin{align}
\nonumber
    W(B) &= \frac{1}{g^{2}} \int \tr (F\wedge F^{*}) + \ln \det M_{0} 
	- \frac{1}{2} \ln \det M_{1} \\
\nonumber
	& + g^{2} \bigl(
\begin{picture}(25,22)(-3,10)
    \qbezier(10,15)(0,29)(10,29)
    \qbezier(10,15)(20,29)(10,29)
    \qbezier(10,15)(0,1)(10,1)
    \qbezier(10,15)(20,1)(10,1)
\end{picture} +
\begin{picture}(40,20)(-10,10)
    \drawline(1,15)(19,15)
    \put(10,15){\ellipse{20}{20}}
\end{picture} +
\begin{picture}(40,20)(-10,10)
    \put(10,15){\arc{20}{0}{0.3}}
    \put(10,15){\arc{20}{0.6}{0.9}}
    \put(10,15){\arc{20}{1.2}{1.5}}
    \put(10,15){\arc{20}{1.8}{2.1}}
    \put(10,15){\arc{20}{2.4}{2.7}}
    \put(10,15){\arc{20}{3.0}{3.3}}
    \put(10,15){\arc{20}{3.6}{3.9}}
    \put(10,15){\arc{20}{4.2}{4.5}}
    \put(10,15){\arc{20}{4.8}{5.1}}
    \put(10,15){\arc{20}{5.4}{5.7}}
    \put(10,15){\arc{20}{6.0}{6.3}}
    \drawline(12,27)(9,25)(12,23)
    \drawline(1,15)(19,15)
\end{picture}
	\bigr) \\
\label{newFI}
	& + \sum_{k=3}^{\infty} g^{2(k-1)} (\text{strongly connected graphs
	    with $k$ loops}).
\end{align}
	From now on we use an ``euclidian trick'' here, changing
    $ x_{0} $ to
    $ i x_{0} $,
	so that
    $ M_{0} $ and
    $ M_{1} $
	become elliptic operators.

	This answer can be considered as an alternative definition
	of the functional integral
(\ref{genfun}).
	Two natural questions can be asked:
	1. are the individual terms in
(\ref{newFI})
	well defined?;
	2. does the series converge?
	Whereas we know almost nothing about the second question, 
	the answer to the first one is quite instructive.
	Here we are confronted with the problem of divergences
	and renormalization.

	Let us turn to the zero order in 
    $ g^{2} $
	term in
(\ref{newFI}).
	It is given by determinants of operators
    $ M_{1} $ and
    $ M_{0} $,
	which clearly diverge and must be regularized.
	The trivial regularization is the subtraction of an infinite
	constant, corresponding for the dets for
    $ B=0 $.
	Then we can use the formula
\begin{equation}
\label{lndet}
    \ln \det M_{i} (B) - \ln \det M_{i}(0) = - \int_{0}^{\infty}
	\frac{dt}{t} \Tr \bigl(e^{-M_{i}(B)t} - e^{-M_{i}(0)t}\bigr) ,
	\quad i=0,1 .
\end{equation}
	The Green functions
    $ D_{i}(x,y;t) $
	of the parabolic equations
\begin{equation}
    \frac{\dd D_{i}}{\dd t} + M_{i} D_{i} = 0, \quad
	D_{i} |_{t=0} = I \delta(x-y)
\end{equation}
	has the well known expansion for small
    $ t $
\begin{equation*}
    D(x,y;t) = \frac{1}{4\pi^{2}t^{2}} e^{- \frac{|x-y|^{2}}{4t}}
	(a_{0}(x,y) + t a_{1}(x,y) +t^{2} a_{2}(x,y) + \ldots) ,
\end{equation*}
	where the coefficients
    $ a_{0}, a_{1}, a_{2}, \ldots $
	are functionals of 
    $ B $.
	(Let me remind, that we deal with 4-dimensional space-time.)
	Trace in
(\ref{lndet})
	means
\begin{equation}
\label{Tr}
    \int \tr D(x,x;t) dx .
\end{equation}
	The coefficient
    $ a_{0} $
	is the holonomy for connection
    $ B $
	along the straight line, connecting points
    $ x $
	and
    $ y $.
	Clearly
    $ a_{0}(x,x) $
	equals unity and so its contribution disappears from
(\ref{lndet})
	due to the subtraction of 
    $ \exp - M(0) t $.
	Now
    $ a_{1}(x,x) $
	for the operator
    $ M_{0} $
	vanishes and same is true for
    $ \tr a_{1}(x,x) $
	for
    $ M_{1} $.
	So what remains is the contribution of 
    $ a_{2} $
	to
(\ref{lndet})
	which diverges logarithmically in the vicinity of 
    $ t=0 $.
	The expansion is valid for small 
    $ t $, 
	so we divide the integration in
(\ref{lndet}) as
\begin{equation}
    \int_{0}^{\infty} = \int_{0}^{\mu} + \int_{\mu}^{\infty}
\end{equation}
	and regularize the first integral as
\begin{multline*}
    \int_{\epsilon}^{\mu} \frac{dt}{t} \int \tr a_{2}(x,x) dx + \\
	+ \int_{0}^{\mu} \frac{dt}{t} \int \bigl(\tr D(x,x;t)
	    - \tr D(x,x;t)|_{B=0} - \tr a_{2}(x,x) 
	      + \mathcal{O}(t^{2})\bigr) dx .
\end{multline*}
	In this way we explicitly separated the infinite part
	proportional to
    $ \ln \epsilon /\mu $.
(In physical literature one uses large momentum cutoff
    $ \Lambda $
instead of short auxilliary time
    $ \epsilon $;
the
    $ \ln \epsilon/\mu $
looks like
    $ -2 \ln \Lambda/m $,
where
    $ m $
has dimension of mass.)

	Now observe, that
    $ \int \tr a_{2}(x,x) dx $
	is proportional to the classical action
    $ \int \tr (F \wedge F^{*}) $.
	It follows from general considerations of gauge invariance
	and dimensionlessness, but can be found also explicitly together 
	with the corresponding numerical coefficient.
	We get
\begin{equation}
\begin{split}
    W(B) = \frac{1}{4}\bigl(\frac{1}{g^{2}} + 
	    \frac{11}{48 \pi^{2}} C(G) \ln \frac{\epsilon}{\mu} \bigr) 
		\int \tr (F\wedge F^{*}) \\
    + \text{finite zero order terms} + \text{higher order loops} .
\end{split}
\end{equation}
	Here
    $ C(G) $
	is a value of a Casimir operator for group
    $ G $
	in the adjoint representation.

	Now we invoke the idea of renormalization 
	a-la Landau and Wilson:
	the coupling constant
    $ g^{2} $
	is considered to be a function of regularizing parameter
    $ \epsilon $
	in such a way that coefficient in front of the classical 
	action stay finite when
    $ \epsilon \to 0 $
\begin{equation}
\label{ren}
    \frac{1}{g^{2}(\epsilon)} + \beta \ln \frac{\epsilon}{\mu}
	= \frac{1}{g_{\text{ren}}^{2}} , \quad
	\beta = \frac{11}{3} \, \frac{C}{16 \pi^{2}} .
\end{equation}
	This can be realized only if the coefficient
    $ \beta $
	is positive, which is true in the case of the Yang-Mills theory.
	Of course
    $ g^{2}(\epsilon) \to 0 $
	in this limit.

	Similar investigation can be done for the quantum equation of motion
(\ref{em}),
	the one loop diagrams are divergent, but the infinite term
	is proportional to the classical equation of motion,
	so that
(\ref{em})
	acquires the form
\begin{equation*}
    \nabla_{\mu} F_{\mu \nu} + g_{\text{ren}}^{2} (\text{finite terms}) = 0 .
\end{equation*}
	Higher loops contribute corrections to the renormalization
(\ref{ren}),
	however their influence is not too drastic.
	I can not explain this here and mention only, that it is due to
	important general statement, according to which the 
	logarithmic derivative of
    $ g^{2}(\epsilon) $
	over
    $ \epsilon $
	does not depend on
    $ \epsilon $
	explicitly
\begin{equation*}
    \frac{\dd g^{2}(\epsilon)}{\dd \ln \epsilon} 
	= \beta \bigl(g^{2}(\epsilon) \bigr),
\end{equation*}
	where
\begin{equation*}
    \beta(g) = \beta g^{3} + \mathcal{O}(g^{5}).
\end{equation*}
	This relation is called the renormalization group
	equation; it follows from it and requirement, that
	the renormalized charge does not depend on
    $ \epsilon $,
	that the correction to
(\ref{ren})
	have form
    $ \ln \ln \epsilon / \mu $
	and lower.

	We stop here the exposition of the elements of quantum field 
	theory and return to our main question of mass.
	We have seen, that important feature of the definition of
    $ W(B) $
	and equations of motion was the appearance of the
	dimensional parameter
    $ \mu $.
	Thus the asymptotic states, which characterize the particle 
	spectrum, depend on this parameter and can be associated
	with massive particles.
	Let us stress, that the divergences are
	indispencible for this,
	they lead to breaking of the scale invariance
	of classical theory.

	In our reasoning it was very important, that divergences have 
	logarithmic character, which is true only for the 4-dimensional 
	space-time.
	All this and positivity of the coefficients
    $ \beta $
	in
(\ref{ren})
	distinguishes the Yang-Mills theory as a unique quantum field
	theory, which has chance to be mathematically correct.

	It is worth to mention, that the disenchantment in quantum
	field theory in the late 50-ties and 60-ties of last century
	was connected with the problem of the charge renormalization.
	In the expressions, similar to
(\ref{ren}),
	for all examples, fashionable at that time, the coefficient
    $ \beta $
	was negative.
	It was especially stressed by Landau after investigation
	of the most successful example of quantum field theory ---
	quantum electrodynamics.
	The realization in the beginning of 70-ties of the fact,
	that in Yang-Mills theory the coefficient
    $ \beta $
	is positive, which is due to 't Hooft, Gross, Wilchek and Politzer,
	changed the attitude of physicists towards the quantum theory
	and led to the formulation of Quantum Chromodynamics.
	(This dramatic history can be found in
\cite{Gross}.)

\section*{Conclusion}
	We have seen, that the quantization of the Yang-Mills field 
	theory leads to a new feature, which is absent in the classical
	case.
	This feature ---
	``dimensional transmutation'' --- is the trading of the 
	dimensionless parameter
    $ g^{2} $
	for the dimensional one
    $ \mu $
	with dimension
    $ [\text{L}]^{2} $.
	Also we have seen, that on a certain level of rigour, 
	the quantization procedure is consistent.
	This gives us hope, that the Clay problem is soluble.
	Of course, the real work begins only now.
	I believe, that the promising direction is the investigation
	of the quantum equation of motion, which should enable to find
	solutions with nontrivial mass.
	One possibility will be a search for solitonic solutions.
	Some preliminary formulas in this direction can be found in
\cite{FN}.

	I hope, that this text could be stimulating for a mathematician
	seriously interested in an actual problem of the modern
	theoretical physics.

\end{document}